\documentclass[10pt,twocolumn,english,nofootinbib,twocolumn,aps,pre]{revtex4-1}
\usepackage{ae,aecompl}
\usepackage[T1]{fontenc}
\usepackage[latin9]{inputenc}
\usepackage{bm}
\usepackage{amsmath}
\usepackage{amssymb}
\usepackage{amscd}
\usepackage{graphicx}
\usepackage{esint}
\usepackage{array}
\usepackage{booktabs }
\usepackage{topcapt}
\usepackage{enumitem}

\usepackage[unicode,breaklinks]{hyperref}
\hypersetup{
    unicode=true,
    plainpages=false, 
    colorlinks=true,
    linkcolor=blue,
    citecolor=blue,
    filecolor=black,
    urlcolor=blue
}

\usepackage{bbold}
\usepackage{calrsfs}

\usepackage{cases}
\usepackage{multirow}

\usepackage{epstopdf}

\date{\today}
\usepackage{babel}
\begin{document}

\title{Efficient and accurate methods for solving the time-dependent spin-1 Gross-Pitaevskii equation}

\author{L. M. Symes$^1$}
\author{R. I. McLachlan$^2$}
\author{P. B. Blakie$^1$}
\affiliation{$^1$Quantum Science Otago, Dodd-Walls Centre for Photonic and Quantum Technologies, Department of Physics, University
of Otago, Dunedin, New Zealand}
\affiliation{$^2$Institute of Fundamental Sciences, Massey University, New Zealand}

\begin{abstract}
We develop a numerical method for solving the spin-1 Gross-Pitaevskii equation. The basis of our work is a  two-way splitting of the spin-1 evolution equation that leads to two exactly solvable flows. We use this to implement a second-order and a fourth-order symplectic integration method. These are the first fully symplectic methods for evolving spin-1 condensates. We develop two non-trivial numerical tests to compare our methods against two other approaches.
\end{abstract}
\maketitle 
 
\section{Introduction} 

A spinor Bose-Einstein condensate is a matter-wave that can access internal spin degrees of freedom \cite{Ho1998a,Ohmi1998a,Stenger1998a}. Theoretically this type of condensate is described by a multi-component (spinor) wavefunction that evolves according to a nonlinear Schr\"{o}dinger equation (NLSE) known as the spinor Gross-Pitaevskii equation (GPE). This system is unique in that it is accurately described by a well-characterized microscopic theory, and it exhibits the phenomena of superfluidity enriched by various types of magnetic (i.e.~spin) ordering. For these reasons spinor condensates are an active topic of research in theory and experiment (e.g.~see \cite{Kawaguchi2012R,Stamper-Kurn2013R}).

We consider the case of a spin-1 system in which the atoms have three internal spin states conveniently labelled by the $z$-projection quantum number as $m=-1,0,+1$. This system has been realized in experiments with ultra-cold $^{87}$Rb and $^{23}$Na atoms. In weak magnetic fields the atomic collisions are rotationally invariant and the nonlinear interactions only depend on two parameters: the spin-independent (i.e.~density) interaction strength $c_0$ and the spin-dependent (i.e.~spin-density) interaction strength $c_1$ (see \cite{Kawaguchi2012R}). For the atoms discussed above $0 < |c_1|\ll c_0$. When $c_1=0$ we get the three-component vector NLSE which is completely integrable in time [for systems that are one-dimensional (1D)] by the inverse scattering method \cite{ablowitz2004discrete}.
When $c_0 = c_1 = -c$, the system is equivalent to a matrix NLSE, which is also completely integrable in 1D by the inverse scattering method \cite{ieda2004exact}.
For general $c_{0,1} \neq 0$ the system is not known to be integrable. Since this is the parameter regime currently accessible to experiments, and since experiments are often in quasi-two-dimensional or fully three-dimensional regimes, it is essential to have numerical methods for solving the system evolution. Previous work in this direction includes an algorithm presented by Wang \cite{Wang2007a} and Bao \textit{et al.}~\cite{Bao2009a} which splits the evolution equation into three parts (flows), which are individually solved, although one of the nonlinear flows is treated approximately. More recently a scheme was proposed which splits the evolution equation into two flows, but uses a brute force diagonalization to construct the flow and a first-order composition method to generate an approximate full solution.

In this paper we develop two new numerical methods for solving the spin-1 Gross-Pitaevskii equation. The basis of our work is a two-way splitting of the spin-1 evolution equation that we show leads to two exactly solvable flows.
We discuss how our splitting can be used to construct second-order and fourth-order symplectic integration algorithms. We develop two non-trivial numerical tests to compare these schemes against two other approaches, and show that our schemes perform well.  

We briefly outline the paper. In Sec.~\ref{Secspin1GPE} we introduce the spin-1 GPE and  our notation. We then discuss how this evolution equation can be spit into two flows in Sec.~\ref{SecSplittings}, and present the analytic solutions to these seperate flows. In Sec.~\ref{SecSympletic} we briefly review symplectic composition and discuss the second-order and fourth-order compositions we use in this paper to realize two algorithms from our flow solutions. In Sec.~\ref{SecApplications}  we discuss some details of a 1D implementation of our algorithm. We then introduce two non-trivial spin-1 problems (a continuous wave solution and the collision of two quasi-solitons) that we use to test the performance of our algorithms against the method developed in Refs.~\cite{Wang2007a,Bao2009a} and against the spin-1 extension of a widely used technique based on the fourth-order Runge-Kutta method. We conclude in Sec.~\ref{Sec:Conclusion}.

\section{Spin-1 GPE}\label{Secspin1GPE}
\subsection{Spin-1  energy functional}
A spin-1 condensate system is described by the spinor field $\bm{\psi} = (\psi_+, \psi_0, \psi_-)^T$, which denotes the condensate field in the $m_F=+1,0,-1$ hyperfine sub-levels respectively.
The ground state energy functional  is \begin{align}
E[\bm{\psi}] &= 
\int d\bm{r}
\left [
\bm{\psi}^\dagger h_0 \bm{\psi} +
\frac{c_0}{2} n^2 + \frac{c_1}{2}|\bm{F}|^2
\right],\label{spinE}
\end{align}
where the single particle hamiltonian is
\begin{align}
(h_0)_{m m^\prime} &= \left[
-\frac{\hbar^2 \nabla^2}{2M} + V_\text{ext} - pm + qm^2
\right] \delta_{m m^\prime},
\end{align}
$V_\text{ext}$ is the external trapping potential (often harmonic). The spatially constant energies  $p$ and $q$ denote the linear and quadratic Zeeman terms, respectively, and arise in experiment from a uniform magnetic bias field along $z$. The nonlinear terms are the total density 
\begin{align}
n = \sum_m |\psi_m|^2,
\end{align}
 and the magnitude of the spin density $\bm{F}=(F_x, F_y, F_z)$, where 
 \begin{align}
 F_\alpha = \bm{\psi}^\dagger f_\alpha \bm{\psi},\quad \alpha \in \{x,y,z\} \label{EqFalpha}
 \end{align}
  with $f_\alpha$ the spin-1 matrices. In particular, we have 
  \begin{align}
  F_z &= |\psi_+|^2 - |\psi_-|^2,\label{Fz}\\
  F_\perp &\equiv F_x + i F_y = \sqrt{2}(\psi_+^* \psi_0 +\psi_0^* \psi_-),\label{Fp}
  \end{align}
   with which we can write $|\bm{F}|^2 = F_z^2 + |F_\perp|^2$. The constants $c_0=\frac{4\pi}{3}(a_0+2a_2)\hbar^2/M$, $c_1=\frac{4\pi}{3}(a_2-a_0)\hbar^2/M$ are the density and spin-dependent interactions, with $a_S$ the $s$-wave scattering length for the spin-$S$ collision channel \cite{Ho1998a}.

\subsection{Spin-1 Gross-Pitaevskii equations}
The time evolution of the condensate field is given by
\begin{align}
i \hbar \frac{\partial \psi_m}{\partial t} = \frac{\delta E}{\delta \psi^*_m}.\label{spinGPEa}
\end{align}
We choose convenient computational units of space $x_0$ (which would usually be set by the trap $V_{\mathrm{ext}}$) and time $t_0=Mx_0^2/\hbar$, and from hereon take all parameters to be expressed in terms of these units.  Then evaluating  (\ref{spinGPEa}) yields the spin-1 GPEs
\begin{align} 
\dot{\psi}_{m} &=
-i\left[-\tfrac{1}{2}{\nabla^2} +qm^2+ V_m \right]\psi_m+ I_m ,\label{sGPEm}
\end{align}
where we have set 
\begin{align}
V_m \equiv V_\text{ext} -pm  + c_0 n,
\end{align}
as the spin-independent effective potential plus linear Zeeman, and
\begin{align} 
I_+ &=  c_1(n_ 0 + F_z)\psi_+
+   c_1\psi_0^2 \psi^{*}_-, \label{Ip1}\\
I_0 &= c_1(n - n_0)]\psi_0 + 2c_1  \psi_+ \psi_- \psi_0^*, \\
I_-&=  c_1(n_0 - F_z)\psi_-
+  c_1\psi_0^2 \psi^{*}_+.\label{Im1}
\end{align}
are the spin-dependent interaction terms, with $n_0=|\psi_0|^2$.

\subsection{Conserved quantities}\label{SecConservQ}
Under evolution according to the spin-1 GPE the condensate has several important conserved quantities, notably energy $E[\bm{\psi}]$ given by Eq.~(\ref{spinE}), normalization
\begin{align}
N[\bm{\psi}] = \sum_m\int d\bm{r}|\psi_m|^2=\int d\bm{r}\, n,
\end{align}
and the $z$-component of magnetization
\begin{align}
M_z[\bm{\psi}] = \int d\bm{r}\left(|\psi_+|^2-|\psi_-|^2\right)=\int d\bm{r}\, F_z.
\end{align}

\section{Splittings}\label{SecSplittings}
We denote the spin-1 GPE as $\dot{\bm{\psi}}=\hat{f}\bm{\psi}$, where $\hat{f}$ denotes the nonlinear operator that actions Eq.~(\ref{sGPEm}), with formal solution $\bm{\psi}(t)=\exp(\hat{f}t)\bm{\psi}(0)$.  For numerical solution it is desirable to break up the operator $\hat{f}$ into two parts as
\begin{align}
\hat{f}=\hat{f}_A+\hat{f}_B.
\end{align}
In breaking up the problem this way we can deal with the individual \emph{flows}  separately
\begin{align}
\dot{\bm{\psi}}=\hat{f}_\mu\bm{\psi},\qquad\mu=A,B,
\end{align}
 and from these compose a solution to the full problem.
 The important feature of the scheme we present here is that we break up the problem into just two flows, which simplifies the composition to higher order schemes, and that for each flow we find an exact solution allowing us to efficiently construct 
\begin{align}
\bm{\psi}(t)=e^{\hat{f}_\mu t}\bm{\psi}(0),\qquad\mu=A,B.\label{fAB}
\end{align}
In constructing these exact flows we consider a spatial domain with periodic boundary conditions.

\subsection{Kinetic and quadratic Zeeman flow ($\hat{f}_A$)} \label{secfA}
We take the first flow to include the kinetic energy and quadratic Zeeman term, i.e.~we define $\dot{\bm{\psi}}=\hat{f}_A\bm{\psi}$ to be
\begin{align}
\dot{\psi}_m=-i\left[-\tfrac{1}{2}\nabla^2+qm^2\right] \psi_m.\label{IP}
\end{align}
This flow is diagonal in Fourier space, yielding the solution 
\begin{align}
\psi_m(t) &= \mathcal{F}^{-1}\left\{ e^{-i(\frac{1}{2}k^2 + qm^2)t} \mathcal{F} [\psi_m(0)] \right\},
\end{align} 
where $\mathcal{F}$ denotes a Fourier transform of the appropriate spatial dimension for the problem and $k=|\bm{k}|$, with $\bm{k}$ the Fourier space coordinate.

\subsection{Trap, linear Zeeman and interaction flow ($\hat{f}_B$)}\label{secfB}
The other flow contains the remaining terms, so that  $\dot{\bm{\psi}}=\hat{f}_B\bm{\psi}$ is given by
\begin{align}
\dot{\psi}_m = -i(V_m\psi_m+I_m).\label{fB}
\end{align}
Importantly, evolution according to (\ref{fB}) conserves both $n$ and $F_z$, thus the $V_m$ are constants in time.
This suggests applying the transformation
\begin{align}
\psi_m = \tilde{\psi}_m e^{-i {V}_mt},\label{Vmt}
\end{align}
which gives the system
\begin{align}
\dot{\tilde{\bm{\psi}}} &=
-ic_1
\begin{pmatrix}
F_z & \tfrac{1}{\sqrt{2}} F_\perp^* & 0 \\
\tfrac{1}{\sqrt{2}} F_\perp & 0 & \tfrac{1}{\sqrt{2}} F_\perp^* \\
0 & \tfrac{1}{\sqrt{2}} F_\perp & -F_z \\
\end{pmatrix} \tilde{\bm{\psi}}
\equiv -i c_1 R \tilde{\bm{\psi}},\label{psit}
\end{align}
where $F_z$ and $F_\perp$ are evaluated according to Eqs.~(\ref{Fz}) and (\ref{Fp}), but using the spinor $\tilde{\bm{\psi}}$.
We also observe that $\tilde{\bm{\psi}}$ evolution according to Eq.~(\ref{psit}) also conserves $\tilde{F}_\perp$, and thus the matrix $R$ is constant in time. 
The solution, by exponentiation, is
\begin{align}
\tilde{\bm{\psi}}(t)
&=\left[
\cos \left(c_1 F t\right) \mathbb{1} 
-i \frac{\sin \left(c_1 F t\right)}{F}
R \right] {\bm{\psi}}(0),\label{EqflowBpsitilde}
\end{align}
with $F=|\mathbf{F}|$, and $\mathbb{1}$ representing the identify matrix. We also note that $\tilde{\bm{\psi}}(0)={\bm{\psi}}(0)$. Hence the complete solution to the $\hat{f}_B$ flow  [i.e.~Eq.~(\ref{fAB})] is given by applying the transform (\ref{Vmt}) to the result in Eq.~(\ref{EqflowBpsitilde}). Our analytic solution (\ref{EqflowBpsitilde}) is key to the methods we develop here. As far as we are aware this solution has not appeared in the literature, although the specialization of this result to the case $F_z=0$ was given in Ref.~\cite{Pu1999}.

We note that we could include $q$  in the $\hat{f}_B$ subsystem, which would be necessary if $q$ was position-dependent. However, the solution in this case involves Jacobi elliptic functions and their inverses \cite{zhang2005coherent} which are slow to compute compared with sines and cosines.

\section{Symplectic composition schemes}\label{SecSympletic}
We can approximate the solution to the GPE by composing the two subsystem solutions. Since our system is Hamiltonian, it is desirable to use a symplectic composition method in order to maintain the Hamiltonian geometric structure of phase space in our solution. Any composition of exact flows is a symplectic method. Both $\hat{f}_A$ and $\hat{f}_B$ preserve $N$, $M_z$ and $E$, thus all composition methods will also preserve $N$, $M_z$ and $E$ (up to roundoff error). In what follows we present a second-order symplectic composition and then discuss a fourth-order method.

\subsection{Second-order symplectic method {\bf (S2)}}
The \textit{Leapfrog} composition is a simple second-order composition for advancing the wavefunction by a time step of size $\tau$. This composition takes the form
\begin{align}
\bm{\psi}(\tau)=e^{\hat{f}\tau}\bm{\psi}(0)\approx e^{ \hat{f}_A\frac{\tau}{2}} e^{  \hat{f}_B\tau} e^{ \hat{f}_A\frac{\tau}{2}}\bm{\psi}(0),
\end{align}
where $e^{\hat{f}_At}$ and $e^{\hat{f}_Bt}$ represents the flow solutions presented in Secs.~\ref{secfA} and \ref{secfB}, respectively. We denote this scheme as  S2 in the results we present.

\subsection{Fourth-order symplectic method {\bf (S4)}}\label{sec4th}
It is possible to employ higher-order compositions. Here we consider  a general fourth-order method, due to Blanes and Moan \cite{Blanes2002a,mclachlan2006geometric}. This composition has the form
\begin{align}
\bm{\psi}(\tau)&=e^{\hat{f}\tau}\bm{\psi}(0)\nonumber\\
&\approx e^{ \hat{f}_A a_6 \tau} e^{ \hat{f}_B b_6 \tau} \cdots e^{ \hat{f}_Aa_1 \tau} e^{ \hat{f}_Bb_1 \tau} e^{\hat{f}_Aa_0 \tau}\bm{\psi}(0),
\end{align}
where we give the values of the constant coefficients $\{a_m,b_n\}$, with $m=\{0,\ldots,6\}$ and $n=\{1,\ldots,6\}$ in Appendix \ref{Sec4thcoefs}. We denote this scheme as  S4 in the results we present.

\section{Applications}\label{SecApplications}
\subsection{Algorithms}
\subsubsection{One-dimensional discretization and stability}\label{secDiscretization}
We consider a spinor condensate in one spatial dimension discretized on a mesh of  $M$ points
\begin{equation}
x_j= -\tfrac{L}{2} + j\Delta x,\qquad j=0,\ldots M-1,
\end{equation}
over the interval $[-\tfrac{L}{2},\tfrac{L}{2})$,
with spacing $\Delta x=L/M$. We assume periodic spatial  boundary conditions, allowing us to implement the Fourer transformations required to implement the kinetic energy operator by Fast-Fourier Transform, with a corresponding reciprocal space mesh
 \begin{equation}
k_j= -k_{\max} + j\Delta k,\qquad j=0,\ldots M-1,
\end{equation}
with $\Delta k=2\pi/L$ and $k_{\max}= {\pi}M/L$.
The discretization we have outlined is equivalent to a spectral approach based on planewaves if the rectangular rule is used as the associated quadrature, i.e.~$\int dx f(x)\to\sum_jf(x_j)\,\Delta x$.

We note that splitting the kinetic energy operator apart from the nonlinear potential in the GPE introduces an instability for any composition method \cite{chin2007higher}, which can only be avoided by choosing the time step size subject to
\begin{align}
\Delta t \lesssim t_{\mathrm{stab}} &\equiv \frac{\pi}{E_{max}}, \\
E_{\max} &= \max_j\left|\frac{1}{2} k_{j}^2 + q\right|,
\end{align}
which says that the highest frequency of the kinetic energy and quadratic Zeeman subsystem must be resolvable over each time step. Satisfying this condition is necessary but not sufficient for ensuring global numerical stability.
It is a severe restriction, with high spatial resolution simulations leading to very small time steps for global stability. Often the step size can be much larger if the integration time of interest is shorter than the time required for the instability to grow.

\subsubsection{Comparison algorithms: W2 and RK4}
We compare our two schemes against two alternative approaches.  

The first method is a second-order time-splitting method, that we denote (W2), introduced by Wang in Ref.~\cite{Wang2007a} (also see \cite{Bao2009a,Bao2010a}). This differs from our scheme because it employs a three-way splitting of the evolution equation, and one of the flows cannot be analytically solved and is treated using a second-order Runge-Kutta step (noting that if this flow was exactly solved then this method would be symplectic). 

The second method we consider is a method based on the fixed time step fourth-order Runge-Kutta method which we denote as (RK4). We have implemented this algorithm using our spatial discretization of Sec.~\ref{secDiscretization} and employing the \textit{interaction picture} technique \cite{Dennis2013a} to exponentiate the kinetic energy flow and improve the algorithm performance.  This type of method is quite commonly used in the BEC community and we have made an immediate extension to the spinor case using the discretization of Sec.~\ref{secDiscretization}. Details of this algorithm are given in Appendix \ref{AppRK4}.

We compare the various algorithms in terms of the number of functions evaluations required per time step in Table \ref{Tabsteps}. The S2 and W2 algorithms require the least evaluations per time step. In 1D tests we find that the S2 is roughly 50\% faster than W2 per time step, with this difference largely arising from the much simpler form of the nonlinear term to evaluate in S2. For larger problems, and particularly higher dimensional problems we expect the single round trip FFT required for the S2 algorithm (using FSAL) will be advantageous.

\begin{table}[h]
\renewcommand{\arraystretch}{1.2}
\begin{tabular}{ccc}
	\hline Algorithm & \quad FFT roundtrips (FSAL) & \quad Nonlinear evals \\
	\hline 	S2 & 2(1) & 1 \\ 
	W2 & 2 & 3 \\ 
	RK4 & 4 & 4 \\ 
	S4 & 7(6) & 7 \\ 
	\hline 
\end{tabular}
\caption{Number of FFT roundtrips (i.e.~FFT and inverse FFT) and nonlinear-term evaluations required for each algorithm. The number in brackets indicates the number required per time-step which can be reduced by the first same as last (FSAL) property (i.e.~last stage is evaluated at the same point as the first stage of the next step).}\label{Tabsteps}
\end{table}

\subsection{Continuous-wave comparison}\label{CWsec}
\subsubsection{Analytic solutions}
Recently an analytic theory was developed for a family of continuous-wave solutions to a uniform spin-1 condensate  \cite{Tasgal2013}. This solution has the feature that it allows the chemical potentials and wavevectors of the components to be different, and thus forms a non-trivial test for numerical methods.
\begin{figure}
	\includegraphics[width=3.1in]{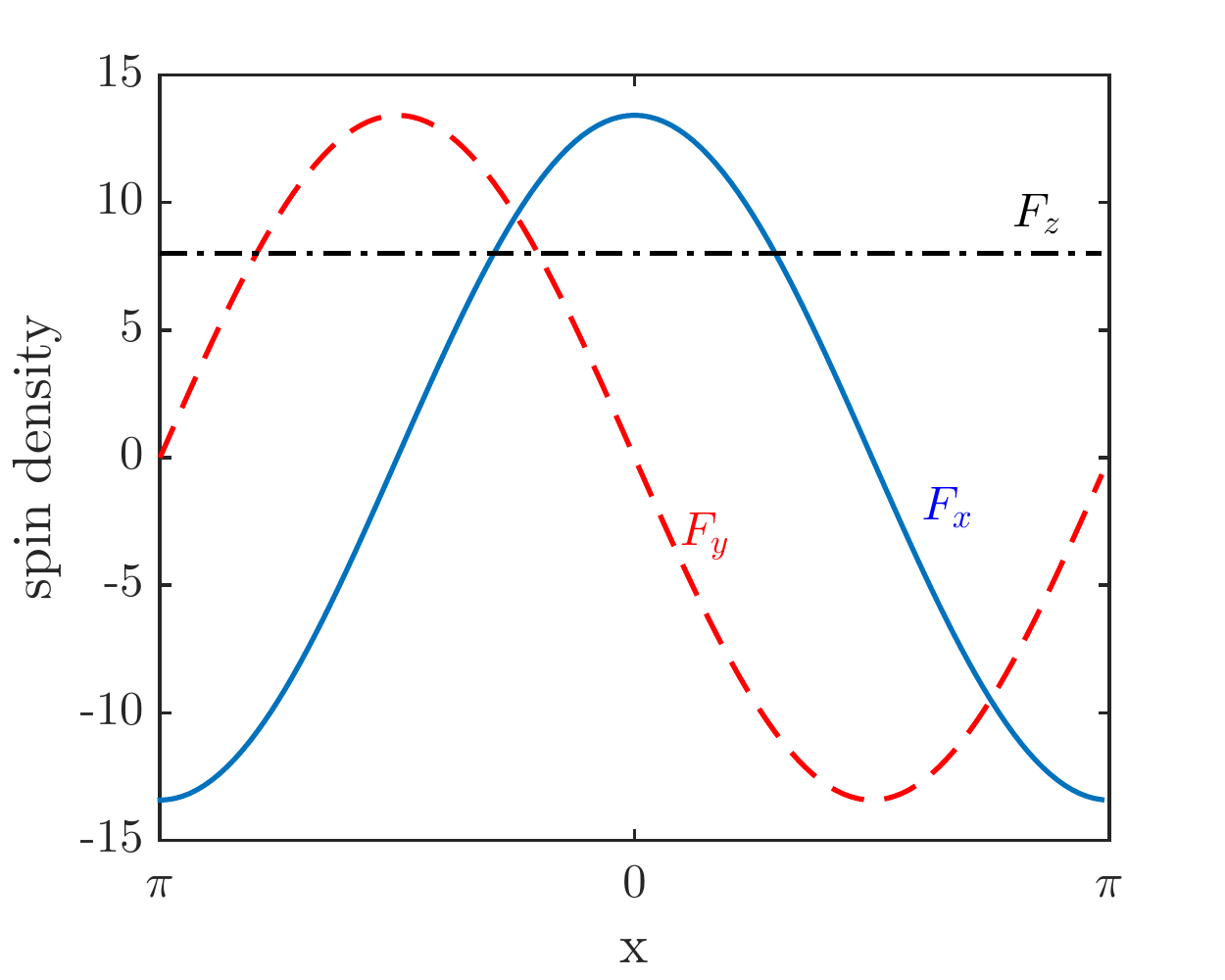}
	\vspace*{-0.1cm}\caption{(Color online) The components of the spin density are shown for a continuous wave solution with parameters: $k_+=5$, $k_-=3$, $\theta_+=0$, $\theta_-=0$, $A_+=3$, $A_-=1$ with $c_0=10$, $c_1=1$ and $q=0.5$. The remaining parameters are determined by the relations given in Appendix \ref{CWsoln}.}\label{FigCWpsi}
\end{figure}

These waves are analytic solutions to the uniform spin-1 GPE for $p=0$ and $c_1\ne0$ and have the form
\begin{align}
\psi_+(x) &= A_+ e^{i( \theta_+ + k_+ x - \omega_+ t)}, \label{cwsolp1}\\
\psi_0(x) &= A_0 e^{i( \theta_0 + k_0 x - \omega_0 t)}, \\
\psi_-(x) &= A_- e^{i( \theta_- + k_- x - \omega_- t)}. \label{cwsolm1}
\end{align}
Here the amplitudes $\{A_-,A_0,A_+\}$ for the three planewave components are taken to be real nonnegative numbers with phases $\{\theta_-,\theta_0,\theta_+\}$. These amplitudes determine the density $n=\sum_mA_m^2$ and magnetization density $F_z=A_+^2-A_-^2$, both of which are conserved quantities (also the individual component densities  $|\psi_m|^2$).
 We have also introduced the component wavevectors  $\{k_-,k_0,k_+\}$ and chemical potentials $\{\omega_-,\omega_0,\omega_+\}$.  When two of the three amplitudes, phases and wavevectors are chosen the remaining values, and the chemical potentials, are determined. There is also an additional parity-like parameter $n_p$ which can take the value of $0$ or $1$, that determines an additional $\pi$ shift in the phase $\theta_0$ relative to $\theta_\pm$. The equations linking these quantities are given in Appendix \ref{CWsoln} (also see \cite{Tasgal2013}). 

An example of this solution is shown in Fig.~\ref{FigCWpsi}, visualized in terms of the components of spin-density [see Eq.~(\ref{EqFalpha})]. 
We see that this solution is a nonlinear spin-wave in $F_\perp$ with the $F_x$ and $F_y$ components of spin density precessing about the constant $F_z$ as the wave propagates.

\subsubsection{Numerical results}
\begin{figure}
	\includegraphics[width=3.1in]{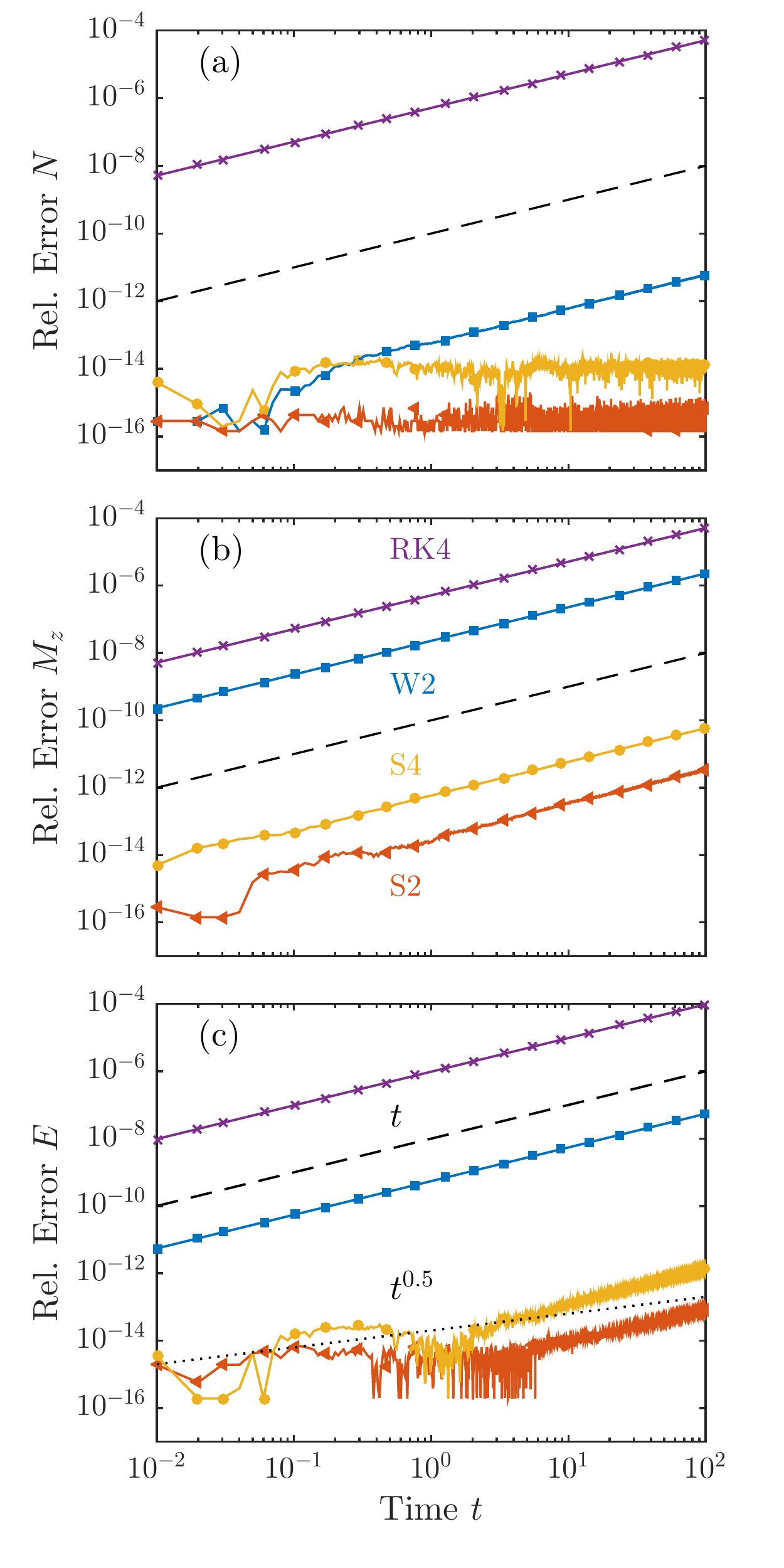}
	\vspace*{-0.5cm}
	\caption{(Color online) Relative errors in the conserved quantities (a) normalization, (b) magnetization and (c) energy for the propagation of the continuous wave solution. The various algorithms, as labeled in (b), all use the same time step ($\tau = 3.125\times10^{-4}$) and  mesh ($M=256$, $L=2\pi$). The initial condition and other simulation parameters are given in Fig.~\ref{FigCWpsi}. Dashed (dotted) line is a guide to the eye to indicate linear (square root) scaling with time $t$.}\label{FigCWrelE}
\end{figure}

 In Fig.~\ref{FigCWrelE} we examine how the various schemes conserve the constants of motion identified in Sec.~\ref{SecConservQ}. Since these quantities should remain constant we characterize the relative error by the absolute relative change in these quantities, i.e.~as
\begin{equation}
\mathrm{Rel.\,Error}\,\, Q\equiv\left|\frac{Q(t)-Q(0)}{Q(0)}\right|,\qquad Q\in\{N,M_z,E\},
\end{equation}
where $Q(t)$ indicates the quantity evaluated at time $t$ during the propagation.
Generally the results show that the symplectic algorithms (i.e.~S2 and S4) and W2 outperform the Runge-Kutta algorithm at conserving the constants of motion, which is expected since symplectic algorithms preserve the geometric properties of phase space.

The S2 and S4 are more accurate than the W2 algorithm.
Although S2 and W2 are both second-order, the W2 method only approximately treats the nonlinear spin exchange terms (see Appendix~\ref{AppW2}) and uses a three-way splitting (which, even for exact flows, is generally less accurate) and thus has larger error coefficients than the S2 method.

We note that W2 and RK4 show linear growth with time $t$ for all three conserved quantities.
S2 and S4 achieve bounded oscillating errors for $N$ (indeed, S2 oscillates around the discretization limit of $2\times 10^{-16}$), and linear error growth in the long-time limit for $M_z$ and $E$.

Symplectic methods should have bounded $E$ errors, while other conserved quantities can be either bounded or linear in time. Bounded errors should grow in time solely due to roundoff errors which accumulate like a random walk, scaling as $t^{1/2}$, however this can be difficult to achieve in practice due to systematic errors \cite{hairer2008}. Figure~\ref{FigCWrelE}(c) shows that S2 and S4 achieve $t^{1/2}$ error scaling until a linear systematic error begins to dominate after $t\sim1$.

One source of systematic error for our methods comes from our implementation of the $\hat{f}_A$ flow not being exact to numerical precision. Of course given a continuous wave our Fourier-space solution is calculated using an exact quadrature, and for other waves achieves exponential accuracy, i.e. it approaches roundoff accuracy in the limit of a large number of discretization points $M$ (although this is quickly offset by increased floating point arithmetic errors from FFTs). In this test the $\hat{f}_A$ flow is accurate to $\sim10^{-15}$. We note that this error is asymmetric across the three components of the wavefunction due to the asymmetry of the wavevectors $k_m$, which means the observed error behaviour is difficult to analyze theoretically.

\begin{figure}
	\includegraphics[width=3.1in]{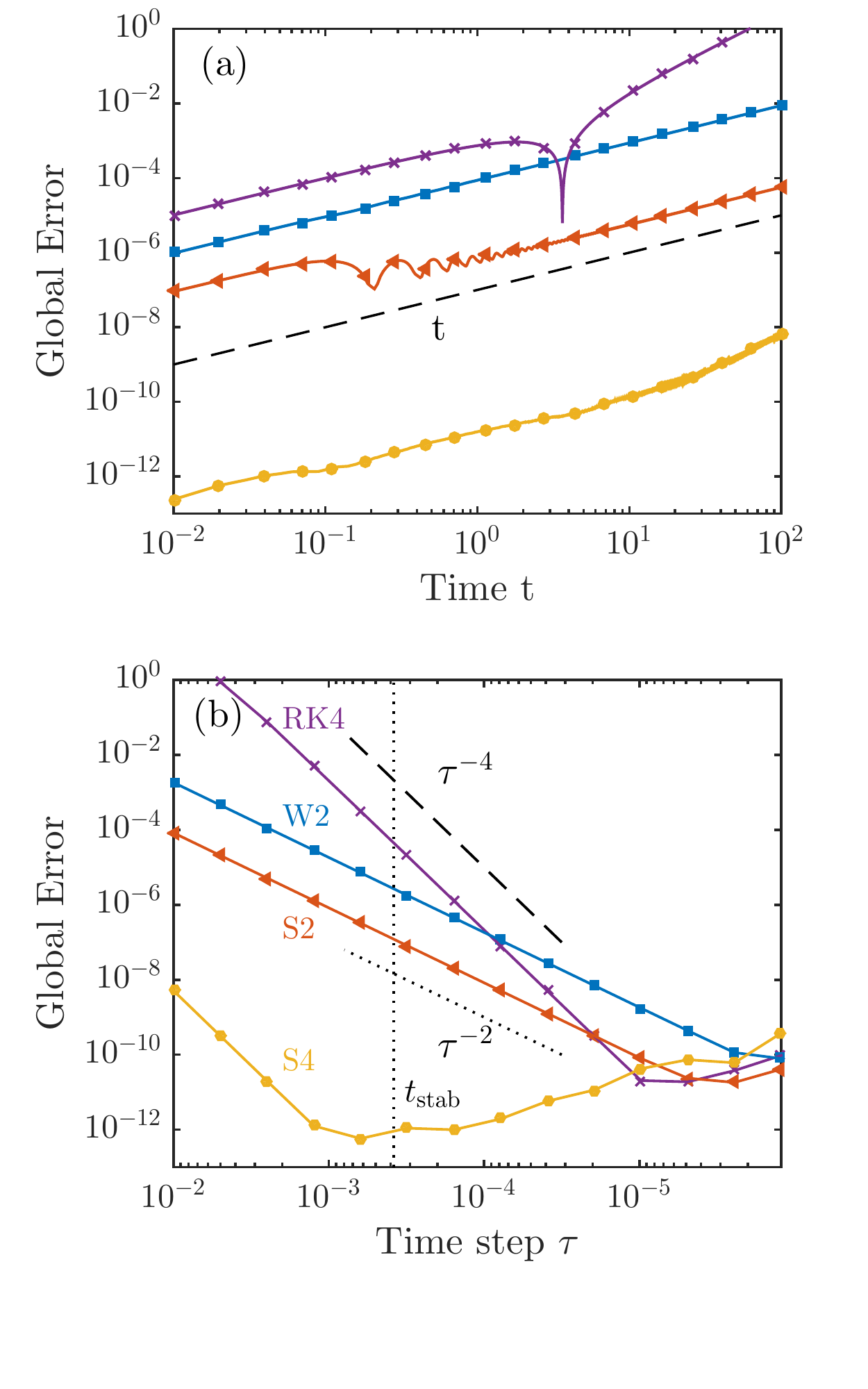}
	\vspace*{-1.2cm}\caption{(Color online) Global error in propagating the continuous wave solution. (a) Global error during a propagation using a time step of $\tau = 6.25\times10^{-4}$.  Dashed line is a guide to the eye showing linear scaling with time $t$. (b) Scaling of the global error after propagation to time $t_f=0.1$ as a function of the as function of the time step $\tau$. The line types for the various algorithms are indicated in (b), as is the estimated stable time step size ($t_{\mathrm{stab}} = 3.8\times10^{-4}$). The dashed (dotted) lines in (b) indicate fourth-order (second-order) power law scaling with respect to $\tau$. Other parameters for the solution are as indicated in Fig.~\ref{FigCWrelE}. }\label{FigCWglobE}
\end{figure}

A more complete test of the numerical error being introduced by each method is to directly compare the wavefunction propagated to time $t$ [which we denote as $\bm{\psi}^{\mathrm{num}}(t)$] against the exact analytic solution  $\bm{\psi}(t)$ given in Eqs.~(\ref{cwsolp1})-(\ref{cwsolm1}). We do this by evaluating the maximum global error introduced by each algorithm, which we define as
\begin{equation}
\mathrm{Global\,Error}\,\, \equiv \max_{x_j,m}\left|{\psi}_m^{\mathrm{num}}(x_j,t)-{\psi}_m(x_j,t)\right|.
\end{equation}
 
 In Fig.~\ref{FigCWglobE}(a) we show the growth of the global error over time for a propagation of fixed time step $\tau=6.25\times10^{-4}$. This is the same propagation examined in Fig.~\ref{FigCWrelE}. The global error shows that S4 is the best at approximating the exact analytic solution by a significant margin, with S2 better than W2, and all of these are better than RK4. Global error grows linearly for all methods until $t\sim1$ [i.e.~the time when the energy error starts growing linearly  for S4 in Fig.~\ref{FigCWrelE}(c)]. At this point both fourth-order methods shift their global error behaviour to be quadratic in time.

 In Fig.~\ref{FigCWglobE}(b) we propogate our numerical solution for a fixed amount of time and consider the scaling of the global error with different time steps $\tau$. These results confirm the expected second-order scaling of the global error with $\tau$ for S2 and W2, as well as the expected fourth-order scaling for RK4 and S4.  For this case the prefactor on the global error for RK4 is up to 9 orders of magnitude larger than S4 depending on the choice of $\tau$. With the S4 method the improvement in global error saturates with decreasing $\tau$ at $\tau\sim10^{-3}$, slightly above the stability threshold, while the other methods show improvement in global error down to $\tau \sim 10^{-5}$. This is because S4 has the most function evaluations per time step and thus accumulates more roundoff errors than the other methods.
We note that due to our choice of a short propagation time, the splitting instability for timesteps greater than $t_\mathrm{stab}$ does not have time to manifest and we observe the expected scaling for all the methods even when $\tau>t_{\mathrm{stab}}$.
 
\subsection{Quasi-soliton comparison}
\subsubsection{Analytic form of spinor quasi-solitons}
\begin{figure}
	\includegraphics[width=3.1in]{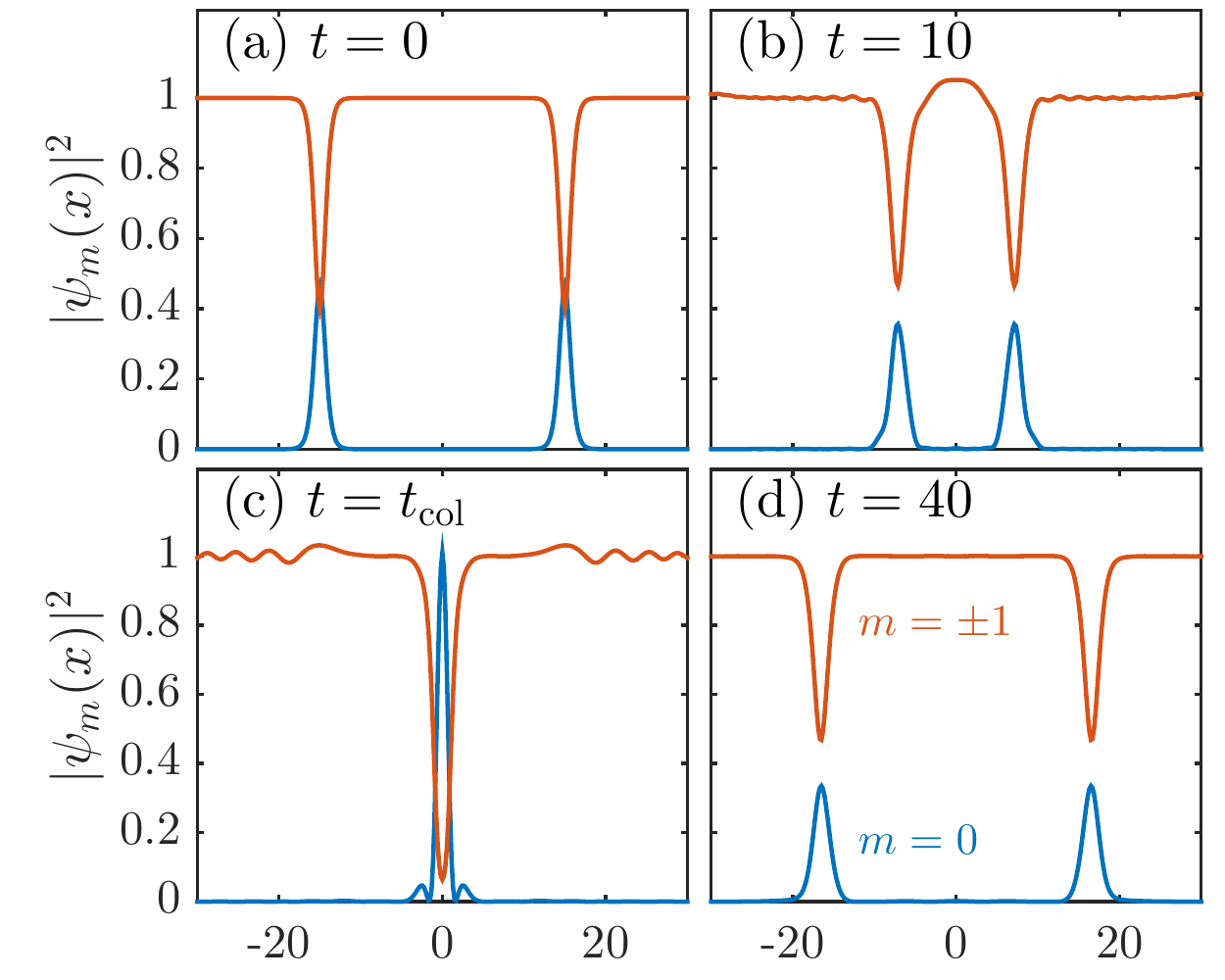}
	\vspace*{-0.5cm}
	\caption{(Color online) Quasi-soliton collision: (a)-(d) snapshots of dynamics showing component densities during the collision dynamics at a range of times. We define the collision time as $t_\mathrm{col}=19$. These results were calculated using S4 with $\tau=0.01$ and match the results presented in Fig.~1 of  Ref.~\cite{Nistazakis2008}. Other parameters: $L=384$, $M= 2048$, $x_0=1$, $c_1=0.314$, $q=0$, $\eta=3.091$, $\xi=1.54$, $\mu=2$. }
	\label{FigQSpsi}
\end{figure}

In the experimentally relevant limit of small spin-dependent interaction, i.e.~$0 < |c_1| \ll c_0$, the spin-1 system is asymptotically equivalent to the Yajima-Oikawa (YO) system, which is integrable by inverse scattering and thus has soliton solutions \cite{Nistazakis2008}. We simulate the collision of two such quasi-solitonic solutions using the initial conditions [given by Eqs.~(37)-(38) of Ref.~\cite{Nistazakis2008}] 
\begin{align}
\psi_\pm(x) &=
\sqrt{\frac{\mu}{2} - \frac{\nu}{2}[\mathrm{sech}^2(x_+) + \mathrm{sech}^2(x_-)]} \\
& \quad \times \exp\left(-i \sqrt{\frac{\nu}{\mu}} [\tanh(x_+) - \tanh(x_-)] \right), \nonumber
\end{align}
\begin{align}
\psi_0(x) &=
\nu^{3/4} \sqrt{\frac{\xi}{\eta \sqrt{\mu}}}
\Big[
\mathrm{sech}(x_+) e^{+i\sqrt{\mu/\nu} x_+ - i (\xi/\nu) x_+}
\\
& \quad
+
\mathrm{sech}(x_-) e^{-i\sqrt{\mu/\nu} x_- + i (\xi/\nu) x_-}
\Big],\nonumber
\end{align}
where $\nu \equiv 4 \eta^2 \delta$, $\delta \equiv c_1/c_0$, $\mu$ is the chemical potential,  $\eta$, $\xi$ are arbitrary parameters, and 
$x_\pm = \sqrt{\nu}(x \pm x_0)$ are the initial positions of the centres of the two pulses. We take the interaction parameter ratio to be  $\delta= 0.0314$ (which is close to the value expected for $^{23}$Na atoms) and choose the other parameters to reproduce the collision studied in Ref.~\cite{Nistazakis2008}. The spinor evolution during the collision is shown in Fig.~\ref{FigQSpsi} (cf.~Fig.~1 of \cite{Nistazakis2008}).
The quasi-solitons are revealed as identical density dips (dark solitons) in the $\pm1$ component densities at the same location that the $m=0$ component has a density peak (bright soliton). For each quasi-soliton, the two dark solitons guide the bright soliton along. We see that under dynamical evolution the two quasi-solitons collide (at $t=19$) and then emerge, roughly keeping their shape.  Because the quasi-soliton is not a true soliton some radiation is emitted during the dynamics, visible as oscillations in the background component densities.  The domain of the simulation $L=384$ is chosen to be much larger than the region of interest containing the solitons in Fig.~\ref{FigQSpsi} to ensure that this radiation does not wrap around (due to periodic boundary conditions) and return to the solitons within the duration of the simulation.

\begin{figure}
	\includegraphics[width=3.1in]{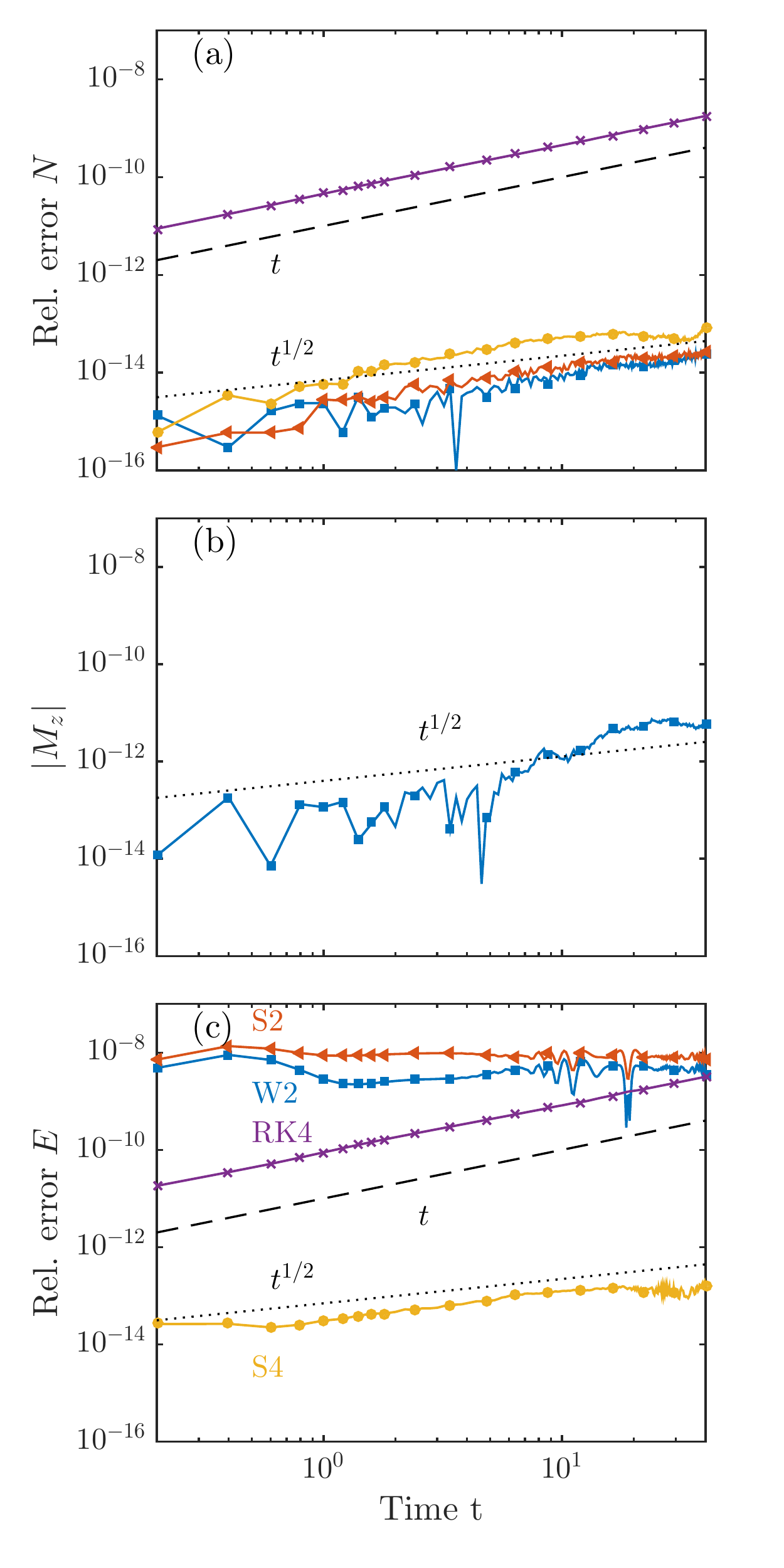}
	\vspace*{-0.5cm}\caption{(Color online) Quasi-soliton evolution: Relative errors in non-zero conserved quantities (a) $N$ and (c) $E$.   (b) shows the magnitude of $M_z$ that develops in the W2 scheme, noting   $|M_z|$ remains zero for the other methods. Dashed (dotted) lines are guides to the eye showing linear (square root) scaling with respect to time $t$. Simulations are for the scenario (initial condition, time and mesh parameters) given in Fig.~\ref{FigQSpsi}.}
	\label{FigQSrelE}
\end{figure}

\begin{figure}
	\includegraphics[width=3.1in]{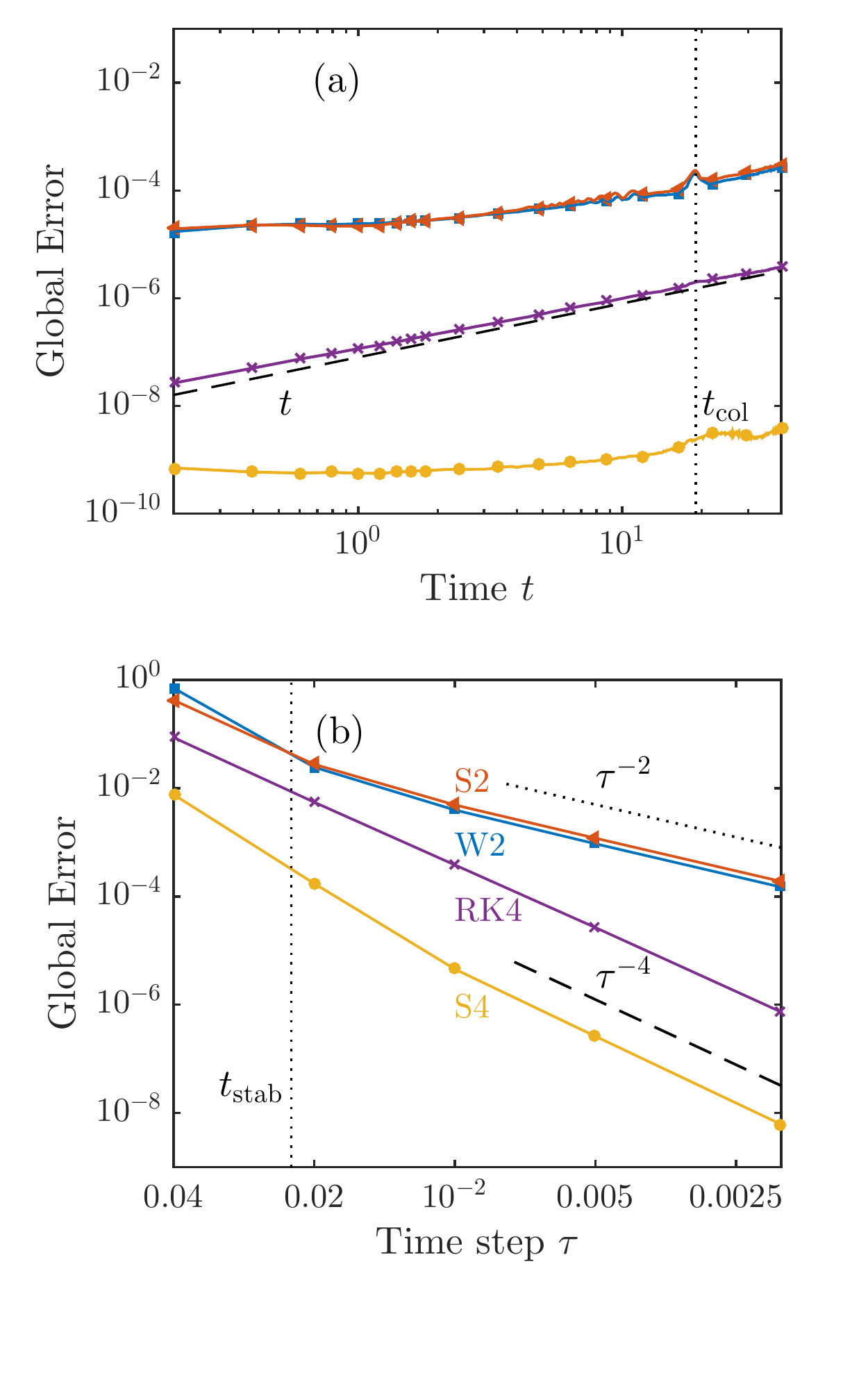}
	\vspace*{-1cm}
	\caption{(Color online) Global error in propagating the quasi-solitons using reference S4 solution with $\tau = 5\times10^{-4}$. (a) Global error during a propagation using a time step of $\tau = 0.02$. Dashed line is a guide to the eye that scales linearly with time $t$, while vertical dash-dotted line shows the time of collision of the quasi-solitons ($t_\mathrm{col}=19$).
		(b) Scaling of the global error with time step $\tau$ of the solitons propagated to $t=2$. Dashed (dotted) line indicates fourth-order (second-order) power law scaling with respect to $\tau$, while vertical dash-dotted line shows the estimated stable time step size ($t_\mathrm{stab}=0.022)$. Simulation parameters and solution mesh are as discussed in  Fig.~\ref{FigQSpsi}.
		}
	\label{FigQSGE}
\end{figure}
\subsubsection{Numerical results}
We compare the the various schemes for the problem shown in Fig.~\ref{FigQSpsi}. First, in Fig.~\ref{FigQSrelE} we compare the relative errors in the conserved quantities $N$ [Fig.~\ref{FigQSrelE}(a)] and $E$ [Fig.~\ref{FigQSrelE}(c)]. 
The symplectic algorithms and W2 are seen to conserve $N$ much better than RK4 at the step size considered in Fig.~\ref{FigQSrelE}(a), with error growth behaviour like $t^{1/2}$ which is indicative of being limited by roundoff. The symplectic algorithms and W2 also show bounded error growth for $E$ in Fig.~\ref{FigQSrelE}(c). In contrast, for RK4 the error in $N$ and $E$ grows linearly with time. The S4 method demonstrates much better conservation of $E$ than all other methods, with $t^{1/2}$ error growth suggesting it has reached a roundoff limit.

The quantity $M_z$ is zero for this soliton (since $|\psi_+|^2=|\psi_-|^2$), thus it is not sensible to use the relative error to compare how well the methods conserve $M_z$. Instead in Fig.~\ref{FigQSrelE}(b) we plot the absolute value of $M_z$ for the W2 method, which is the only scheme that develops a non-zero value.
This is not simply due to roundoff, which would give $t^{1/2}$ scaling, but because the W2 method uses an approximate solution to the nonlinear spin exchange terms which does not explicitly conserve $F_z$ (see Appendix \ref{AppW2}).

Because we do not have an exact analytic solution for the dynamics of the soliton collision we are unable to make the same global error comparison we performed for the continuous wave solutions. However, we use the S4 method with $\tau=5\times10^{-4}$ to produce reference results which we compare the methods against for appreciably larger time steps.
In Fig.~\ref{FigQSGE}(a) we show the scaling of the global error with time $t$. RK4 scales almost linearly while the symplectic methods and W2 slowly increase as they evolve to the collision at $t_\mathrm{col}$.
The results in Fig.~\ref{FigQSGE}(b) demonstrate the expected fourth-order scaling with $\tau$ for S4 and RK4, and the second-order scaling for W2 and S2. S4 is again the most accurate method by several orders of magnitude. In contrast to the Fig.~\ref{FigCWglobE}(b), here RK4 is more accurate than S2 and W2 due to the fourth-order scaling having set in faster.

\section{Conclusions}\label{Sec:Conclusion}
We have demonstrated that a two-way splitting of the spin-1 evolution equation leads to two exactly solvable flows which are convenient for numerical implementation as a symplectic integration scheme. Notably, our approach leads to a simpler and more efficient second-order algorithm compared to the earlier work by Wang \cite{Wang2007a}, and because we are able to make a two-way splitting we can extend the scheme to fourth-order using the composition method of Blanes and Moan \cite{Blanes2002a}. We have presented several tests that show our methods are accurate.
A feature of symplectic methods is that they can provide highly accurate solutions for Hamiltonian systems over long time periods. The development of such methods for the spinor system is timely given the increasing interest in the long-time coarsening dynamics of this system (e.g.~see \cite{Mukerjee2007a,Kudo2013a,Kudo2015a,Williamson2016a}).  
While our tests are based on 1D spatial problems with periodic boundary conditions, our methods are immediately extendible to higher dimensional systems and to  non-periodic spatial domains, e.g.~by using a finite difference or finite element description of $\hat f_A$ combined with a Krylov approximation to $e^{\hat f_A}\psi$ (see \cite{Hochbruck1998a}). 

It is also worth making a few remarks relating the differences between our approach and that of Wang \cite{Wang2007a} (which we outlined in Appendix \ref{AppW2}), and the potential to extend these ideas to more general cases, including higher spin systems. Wang splits up the terms arising for the spin-dependent energy ($\mathbf{F}^2$) into separate flows, while we keep these terms together.
The solution we have developed here utilizes that the quadratic Zeeman term is spatially uniform (a good approximation in many experiments) and can thus be included with the kinetic energy flow. However, our technique can be extended to situations where the quadratic Zeeman energy varies with position. To do this the quadratic Zeeman term needs to be included with the $\hat{f}_B$ flow, which admits an analytic solution (e.g.~see \cite{zhang2005coherent}) in terms of Jacobi elliptic functions.
The extension to higher spin is less immediate and is in need of attention.
Schemes in the literature \cite{Wang2011a,Gawryluk2015a} require explicit numerical diagonalization of the nonlinear spin subsystem at each time-step,
which for the spin-$f$ case consists of $f+1$ interaction terms (e.g.~for $f=2$ a spin singlet-pair term with new interaction coefficient $c_2$ emerges). Like what we have demonstrated here with the spin-1 case, it may be possible through careful analysis and exploiting the symmetry of the system to develop simpler splittings for higher spin systems that are easier and more efficient to implement.

\section*{Acknowledgments}
We acknowledge support by the Marsden Fund of New Zealand.  

\appendix

 \section{Coefficients for fourth-order composition}\label{Sec4thcoefs}
 The $a$-coefficients for the fourth-order composition presented in Sec.~\ref{sec4th} are (see \cite{Blanes2002a,mclachlan2006geometric})
 \begin{align} 
a_0 &=a_6 =0.0792036964311957, \\
a_1 &=a_5 =0.353172906049774, \\
a_2 &=a_4 = -0.0420650803577195, \\
a_3 &=1 - 2(a_0 +a_1 +a_2),
\end{align}
while the $b$-coefficients are
 \begin{align} 
b_1 &=b_6 =0.209515106613362, \\
b_2 &=b_5 = -0.143851773179818, \\
b_3 &= b_4 = 0.5 - (b_1 + b_2).
\end{align}

\section{Wang method (W2)} \label{AppW2}
Here we briefly review the method introduced by Wang \cite{Wang2007a} for solving the spin-1 GPE.
This method  has three stages,
\begin{align}
\psi(t) &= e^{\tfrac{t}{2}C} e^{\tfrac{t}{2}D} e^{t G} e^{\tfrac{t}{2}D} e^{\tfrac{t}{2}C} \psi(0).
\end{align}
The $C$ flow consists of the kinetic energy term,
\begin{align}
\dot{\bm{\psi}} &= i \tfrac{1}{2}\nabla^2 \bm{\psi}, \\
e^{\tau C} \bm{\psi} &= \mathcal{F}^{-1}\left\{ e^{-i \tau \frac{1}{2}k^2} \mathcal{F} [\bm{\psi}] \right\},
\end{align}
where $k=|\bm{k}|$, with $\bm{k}$ the Fourier space coordinate. The $D$ flow has the quadratic Zeeman and density-like terms, and is solved exactly as
\begin{align}
e^{\tau D} \psi_m &= e^{-i \tau [qm^2-pm+ c_0 n + c_1 (n-2|\psi_{-m}|^2)]} \psi_m,
\end{align}
while the $G$ flow has the nonlinear spin-exchange terms,
\begin{align}
\dot{\bm{\psi}} &= -i c_1
\begin{pmatrix}
0 & \psi_-^* \psi_0 & 0 \\
\psi_0^* \psi_- & 0 & \psi_0^* \psi_+ \\
0 & \psi_+^* \psi_0 & 0
\end{pmatrix} \bm{\psi} \equiv G \bm{\psi},
\end{align}
which cannot be solved exactly. Instead an explicit second-order Runge-Kutta method (Heun's method)  is employed
\begin{align}
\int_0^\tau G(\bm{\psi}(t)) d\tau &\approx \tfrac{\tau}{2}(G(\bm{\psi}(0)) + G(\tilde{\bm{\psi}}(\tau)) \equiv \tau R_w, 
\end{align}
where
\begin{align}
\tilde{\bm{\psi}}(\tau) &= \bm{\psi}(0) + \tau G(\bm{\psi}(0)) \bm{\psi}(0).
\end{align}
The G flow is then approximated by
\begin{align}
e^{\tau G} \bm{\psi}(0) &\approx e^{\tau R_w} \bm{\psi}(0),
\end{align}
with an analytic expression for the exponentiated matrix $e^{\tau R_w}$ allowing this to be evaluated accurately (see \cite{Wang2007a} for details).

We note that the $D$ and $G$ flows 
conserve $M_z$ exactly, and each conserve their relevant energy term, the sum of which equals the density plus spin energy. But the Runge-Kutta approximation of the $G$ flow does not conserve $M_z$ or its associated energy exactly. This is expected since explicit Runge-Kutta methods are not symplectic. Due to using an inexact flow, the method is not symplectic; however, numerical results show that it can behave similarly to our symplectic method S2 in particular cases, e.g. see Figs.~(\ref{FigQSrelE}-\ref{FigQSGE}).

\section{Fourth Order Runge-Kutta method (RK4)} \label{AppRK4}
 An alternative method that has been employed to solve the spin-1 GPE involves using the Runge-Kutta method to integrate the interaction terms.  
 We make the so called interaction picture transformation [equivalent to solving the kinetic energy and quadratic Zeeman flow (\ref{IP})] using the unitary transformation
$\bm\phi(t)=U(t) \bm\psi(t)$, which is given (component-wise) by
\begin{align}
\phi_m=e^{+i(t - t^\prime)\left[-\frac{1}{2}\nabla^2 + qm^2\right]}\psi_m,
\end{align}
where $t^\prime$ is the temporal origin of the interaction picture.
This interaction picture field then evolves according to the evolution equation
\begin{align}
\dot{\bm{\phi}} &= U(t) \hat{f}_B U^\dagger(t) \bm{\phi}
\equiv \hat{f}_C(t)\bm{\phi},
\end{align}
where $\hat{f}_B$ is as defined in Sec.~\ref{secfB}. This equation is then solved discretely in time according to the standard Runge-Kutta algorithm

\begin{align}
\bm{\phi}(t_n) &= U(t_n) \bm{\psi}(t_n), \\
\bm{k}_1&=\hat{f}_C(t_n)\bm{\phi}(t_n),\\
\bm{k}_2&=\hat{f}_C(t_n + \tfrac{\tau}{2})\left[\bm{\phi}(t_n)+\tfrac{\tau}{2}\bm{k}_1\right], \label{eq:RK-k2}\\
\bm{k}_3&=\hat{f}_C(t_n + \tfrac{\tau}{2})\left[\bm{\phi}(t_n)+\tfrac{\tau}{2}\bm{k}_2\right], \label{eq:RK-k3}\\
\bm{k}_4&=\hat{f}_C(t_n + \tau)\left[\bm{\phi}(t_n)+\tau\bm{k}_3\right],
\end{align}
yielding
\begin{align}
\bm{\phi}(t_{n+1}) = \bm{\phi}(t_n) + \frac{\tau}{6}(\bm{k}_1+2\bm{k}_2+2\bm{k}_3+\bm{k}_4).
\end{align}
By choosing $t^\prime = t_n + \tfrac{\tau}{2}$, then $\hat{f}_C(t_n + \tfrac{\tau}{2}) = \hat{f}_B$ for Eqs.~(\ref{eq:RK-k2})-(\ref{eq:RK-k3}), and we only need $4$ pairs of Fourier transforms, i.e.~
\begin{align}
\bm{k}_1&=U(t_n) \hat{f}_B \bm{\psi}(t_n),\\
\bm{k}_2&=\hat{f}_B\left[\bm{\phi}(t_n)+\tfrac{\tau}{2}\bm{k}_1\right],\\
\bm{k}_3&=\hat{f}_B\left[\bm{\phi}(t_n)+\tfrac{\tau}{2}\bm{k}_2\right],\\
\tilde{\bm{k}}_4&=\hat{f}_B U^\dagger(t_n + \tau)\left[\bm{\phi}(t_n)+\tau\bm{k}_3\right],
\end{align}
\begin{align}
\bm{\psi}(t_{n+1}) &= U^\dagger(t_n + \tau)\left\{
\bm{\phi}(t_n) +  \frac{\tau}{6}(\bm{k}_1+2\bm{k}_2+2\bm{k}_3)
\right\}
+ \tau \tilde{\bm{k}}_4/6.
\end{align}
We note that
\begin{align}
U(t_n) = U^\dagger(t_n + \tau) &= e^{-i\frac{\tau}{2}\left[-\frac{1}{2}\nabla^2 + qm^2\right]},
\end{align}
thus for fixed time steps $\tau$ we can pre-compute the Fourier space coefficients.

\section{Parameters in the continuous-wave solution}\label{CWsoln}

Here we briefly outline the relationships between the parameters of this continuous wave solution presented in Sec.~\ref{CWsec}, noting that additional details can be found in Ref.~\cite{Tasgal2013}.  

For Eqs.~(\ref{cwsolp1})- (\ref{cwsolm1}) to be a solution of the spin-1 GPE we must have
\begin{align}
k_0 &= \tfrac{1}{2} (k_+ + k_-), \\
\omega_0 &= \tfrac{1}{2} (\omega_+ + \omega_-), \\
\theta_0 &= \tfrac{1}{2} (\theta_+ + \theta_- + n_p\pi),
\end{align}
where $n_p=\{0,1\}$,
and the chemical potentials are given by
\begin{align} 
\omega_\pm &= \tfrac{1}{2} k_\pm^2 + U_\pm + c_1 \left[(-1)^{n_p} A_0^2 - 2A_+ A_-\right] \frac{A_\mp}{A_\pm} , \\
\omega_0 &= \tfrac{1}{2} k_0^2 + U_0 - c_1 \left[A_0^2 - (-1)^{n_p} 2A_+ A_-\right],
\end{align}
where we have set $U_m\equiv (c_0 + c_1)(A_+^2 + A_0^2 + A_-^2)+qm^2$.
From the constraint on $\omega_0$ we find $A_0$ in terms of $A_\pm$
\begin{align}
A_0^2 &= 2(-1)^{n_p} A_+ A_- \left(1 - \frac{\tfrac{1}{2} [(k_+ - k_-)/2]^2 + q}{c_1[A_+ + (-1)^n A_-]^2} \right).
\end{align}

\bibliographystyle{apsrev4-1}

\end{document}